\documentclass[%
reprint,
superscriptaddress,
amsmath,amssymb,longbibliography,
aps,prb
]{revtex4-2}
\usepackage[utf8]{inputenc}
\usepackage[T1]{fontenc} 
\usepackage{graphicx}
\usepackage{dcolumn}
\usepackage{bm}
\usepackage{amsmath}
\usepackage{lmodern}
\usepackage{mathtools}
\usepackage{float}
\usepackage[colorlinks,citecolor=blue,linkcolor=blue,urlcolor=blue]{hyperref}
\usepackage{siunitx}
\begin{document}
\title{Spin~polarization~of~exciton-polariton~condensate in~a~photonic~synthetic~effective~magnetic~field}
\author{R.~Mirek}
\author{M.~Furman}
\author{M.~Kr\'ol}
\author{B.~Seredy\'{n}ski}
\author{K.~Łempicka-Mirek}
\author{K.~Tyszka}
\author{W.~Pacuski}
\affiliation{Institute of Experimental Physics, Faculty of Physics,\\University of Warsaw, ul. Pasteura 5, PL-02-093 Warsaw, Poland}
\author{M.~Matuszewski}
\affiliation{The Institute of Physics, Polish Academy of Sciences,\\ Aleja Lotnik\'ow 32/46, PL-02-668 Warsaw, Poland}
\author{J.~Szczytko}
\author{B.~Pi\k{e}tka}
\email{barbara.pietka@fuw.edu.pl}
\affiliation{Institute of Experimental Physics, Faculty of Physics,\\University of Warsaw, ul. Pasteura 5, PL-02-093 Warsaw, Poland}

\begin{abstract}
We investigate the spin polarization of localized exciton-polariton condensates. We demonstrate the presence of an effective magnetic field leading to the formation of elliptically polarized condensates. We show that this synthetic field has an entirely photonic origin, which we believe is unique for the CdTe-based microcavities. Moreover, the degree of spin polarization of localized polariton condensates in samples with magnetic ions depends on the excitation power or polarization of the non-resonant excitation laser. In an external magnetic field, the semimagnetic condensate spontaneously builds up strong spin polarization. Based on the magnetic field behavior of the condensate in the presence of magnetic ions, we apply a model that allows us to estimate the polariton-polariton interaction strength in a CdTe-system to approx. 0.8\,$\mu \text{eV}\!\cdot\!\mu \text{m}^2$.
	
\end{abstract}
\maketitle
\section{\label{sec:intro}Introduction}

The optoelectronic devices that have emerged recently, explore the potential that light provides over electrons in terms of high speed, long propagation distances, and high operation rates. To implement efficient computation protocols and data processing in photonic nanostructures, the spin state of photons has to be controlled and easily manipulated \cite{Hirohata_JoMaMM2020}. Exciton-polariton non-equilibrium Bose-Einstein condensates (C) have recently appeared as a perfect platform for the realization of photonic devices \cite{Gao_PRB2012, Nguyen_PRL2013, Cerna_NatComm2013, Ballarini_NatComm2013, Anton_PRB2013, Marsault_APL2015, Dreismann_NatMat2016, Zasedatelev_NatPhot2019}. Exciton-polaritons are half-light half-matter quasiparticles in which both components play a significant role \cite{Weisbuch_PRL1992, Savona_SSC1995}. The photonic part allows for easy manipulation of light intensity, phase, and polarization, but also spatial localization and propagation. The excitonic component provides strong nonlinear effects into this photonic platform \cite{Ciuti_PRB1998, Wouters_PRB2007, Schumacher_PRB2007, Vladimirova_PRB2010, Tan_PRX2020, Gu_NatComm2021}. 

However, the lack of application in spin-related polaritonics comes from the limited control of the condensate spin. The spin polarization of exciton-polaritons directly translates into the polarization of the light emitted spontaneously in the recombination process of these quasiparticles \cite{Kavokin_PRL2004}. Polariton spin can be controlled through both: the excitonic and photonic components. 

The excitonic part can be easily manipulated by the external magnetic field, but in many cases it requires a high field intensity \cite{Pietka_PRB2015}. In the absence of a magnetic field, typical condensates composed of nonmagnetic excitons have manifested linear polarization under non-resonant excitation \cite{Klopotowski2006, Shelykh_SST2010, Sala_PRB2016} which led to a condensate with the same occupation of spin-up and spin-down states agreeing with theoretical predictions \cite{Laussy_PRB2006}. It has been possible to create an elliptically polarized condensate at zero magnetic field, but in most cases this effect has been stochastic and difficult to control \cite{Ohadi_PRX2015, Pickup_PRL2018}. The stable and controllable degree of circular polarization has also been observed as a result of the small ellipticity of a non-resonant laser pump \cite{Askitopoulos_PRB2016, Klaas_PRB2019, Gnusov_PRB2020}
. Finally, optical control of the polariton spin using a circularly polarized laser as an excitation pump has been reported in many works \cite{Paraiso_NatMat2010, Ohadi_PRL2012, Anton_PRB2015, Pickup_PRB2021, Real_PRR2021}.

The photonic component is much more difficult to control. Engineering of photonic states is done through photonic lattices, photonic traps, or cavities filled with birefringent media \cite{Tercas_PRL2014, Lim_NatComm2017, Klembt_Nat2018, Whittaker_NatPhot2020, Schneider_RepPP2016, Solnyshkov_TOS2021} where the structuring imposed on the photonic mode creates a synthetic magnetic field.

In this work, we show the effects of a real and synthetic magnetic field acting on polariton condensates in a planar cavity. We tuned the spin properties of exciton-polaritons in microcavity with semimagnetic quantum wells through the excitonic component and magnetic ions. However, in the absence of a real magnetic field, the spin polarization of the condensate was induced by a purely photonic effect. 
This allowed us to create condensates in a close vicinity, having opposite spin polarization within the same excitation beam, which was not possible in previously reported realizations (Fig.~{\ref{fig:fig1}}). This effect is stable in time and space and works as a synthetic effective magnetic field. We discuss in detail the effects of magnetic ions and both real and synthetic magnetic fields.

\section{\label{sec:results} Experimental results}
\subsection{\label{sec:results_A} Intrinsic spin polarization of the condensate}

\begin{figure}
	\centering
	\includegraphics[width=.48\textwidth]{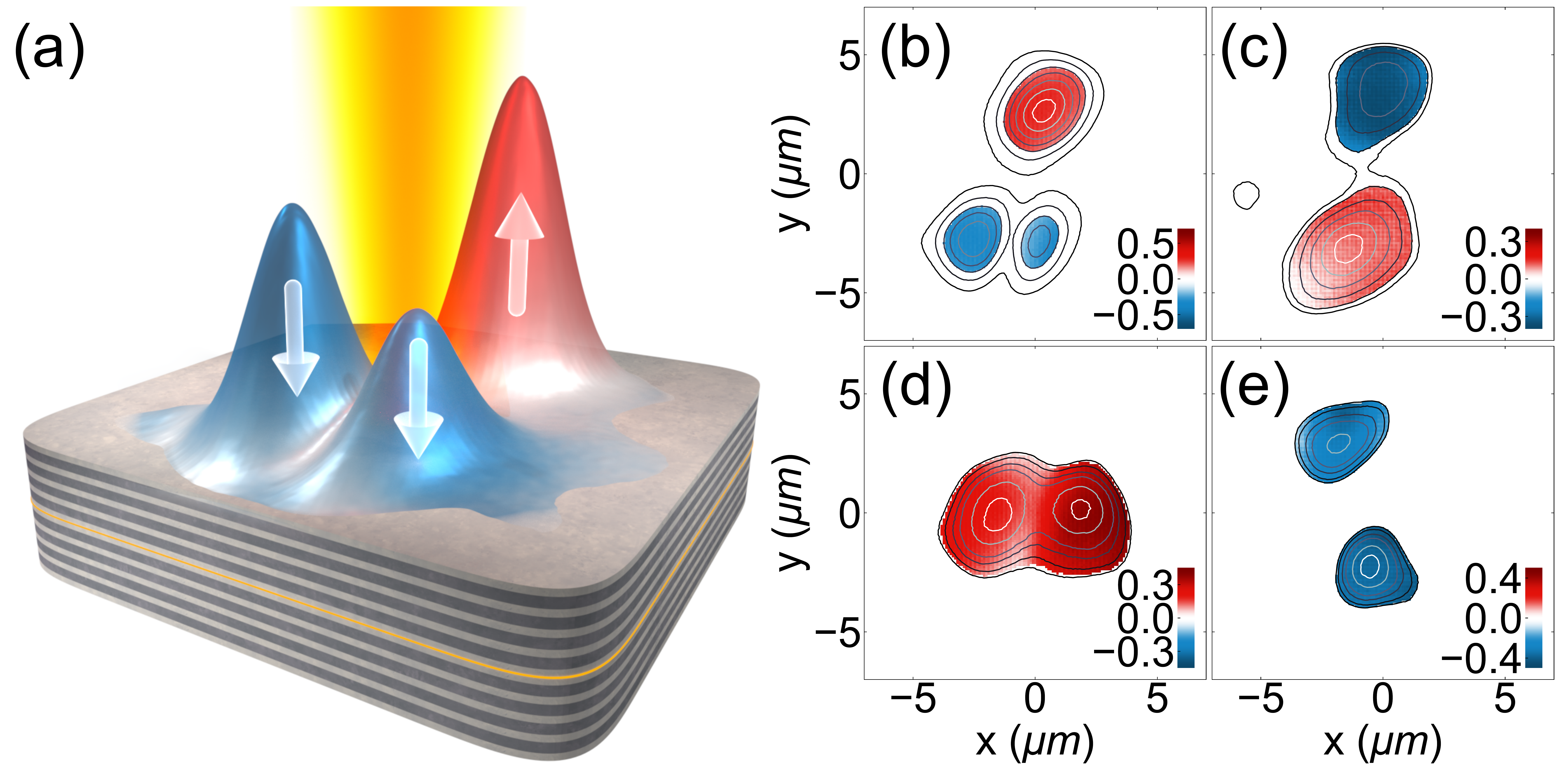}
	\caption{Creation of the elliptically polarized condensates by a linearly polarized laser beam. (a) Schematic illustration of an excitation laser forming three spatially separated condensates with different elliptical polarization. (b)--(e) Degree of circular polarization for various condensate arrangements realised on different positions on the sample. Color scale gives information about the degree of circular polarization with constant intensity contours.}
	\label{fig:fig1}
\end{figure}

 We investigated II-VI microcavity with four \SI{20}{\nano\meter}-thick semimagnetic quantum wells containing 1\% of manganese ions \cite{Rousset_APL2015}. The presence of magnetic ions leads to the \textit{s,p-d} exchange interaction between spins of localized \textit{d}-electrons and spins of delocalized carriers from \textit{s} and \textit{p} orbitals. The use of a semimagnetic sample gives access to the observation of spin-dependent phenomena that are not present in nonmagnetic samples \cite{Ulmer-Tuffigo_JCG1997, Haddad_SSC1999, Mirek_PRB2017, Rousset_PRB2017, Krol_PRB2019}. 
We excited the sample nonresonantly with linearly polarized \SI{4}{\pico\second} laser pulses. The degree of circular polarization of the created localized condensates is presented in Fig.~\hyperref[fig:fig1]{\ref{fig:fig1}(b-e)}. We define the degree of circular polarization $\wp$ as a normalized difference between the light intensities observed in opposite circular polarizations $\wp = \frac{I_{\sigma^+}-I_{\sigma^-}}{I_{\sigma^+}+I_{\sigma^-}}$. The \SI{10}{\micro\meter} diameter laser spot created counter- (b-c) or co- (d-e) polarized condensates simultaneously. All these spatial arrangements of spin-polarized condensates are different, and the configuration depends only on the position on the sample. Different locations have different photonic disorder caused by the growth conditions of CdTe-based materials. The lateral imperfection of the distributed Bragg mirrors and the cavity has two roles: it allows for localization of condensates in two-dimensional potential minima and is responsible for a persistent nonzero circular polarization. As we show later, this latter effect is observed only above the condensation threshold. Such optical activity is unique for II-VI materials and has already been manifested, for example, in the pinning of half-quantum vortices \cite{Lagoudakis_Science2009}. 

\begin{figure}
	\centering
	\includegraphics[width=0.48\textwidth]{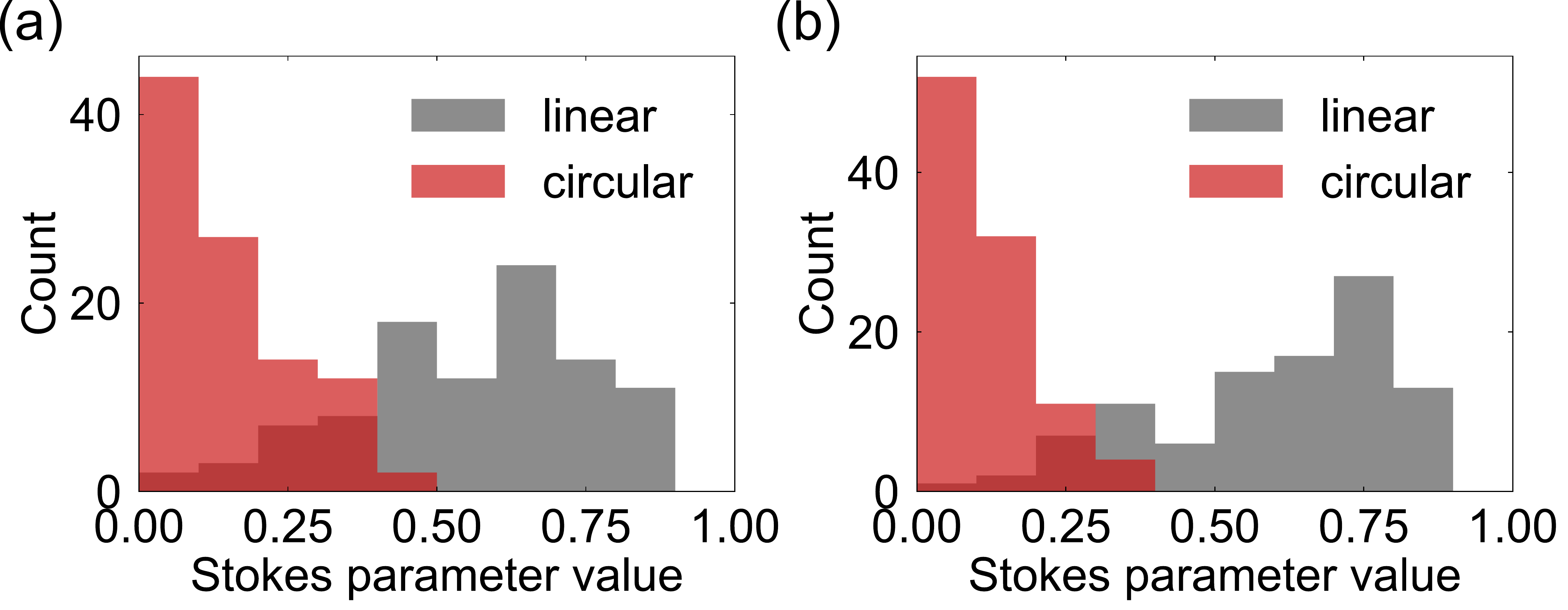}
	\caption{Distribution of a linear and a circular polarization component for randomly chosen condensates investigated in the sample (a) with and (b) without manganese ions in the structure.}
	\label{fig:SI4}
\end{figure}

The spontaneous formation of spin-polarized condensates in the absence of an external magnetic field is a very prominent effect. The degree of circular polarization can reach up to 0.5 under excitation with a linearly polarized non-resonant laser, what can be observed in Fig.~\hyperref[fig:SI4]{\ref{fig:SI4}(a)}. We studied the polarization of one hundred condensates in randomly chosen positions on the sample. Based on the measurements of the Stokes polarization parameters $S_1, S_2$ and $S_3$ we obtained an absolute value of the linear ($\sqrt{S_1^2 + S_2^2}$) and circular ($S_3$) components to the condensate polarization. To prove the importance of the synthetic magnetic field and simultaneously completely exclude the role of manganese ions in the spontaneous build-up of circular polarization of the condensate at zero magnetic field, we studied an analogous sample without manganese ions in a structure [Fig.~\hyperref[fig:SI4]{\ref{fig:SI4}(b)}]. The condensates observed in both samples had random polarization in the absence of magnetic field, and their distribution did not differ qualitatively.

\subsection{\label{sec:results_B} Spin of the condensate tuned by magnetic field}

\begin{figure}
	\centering
	\includegraphics[width=0.48\textwidth]{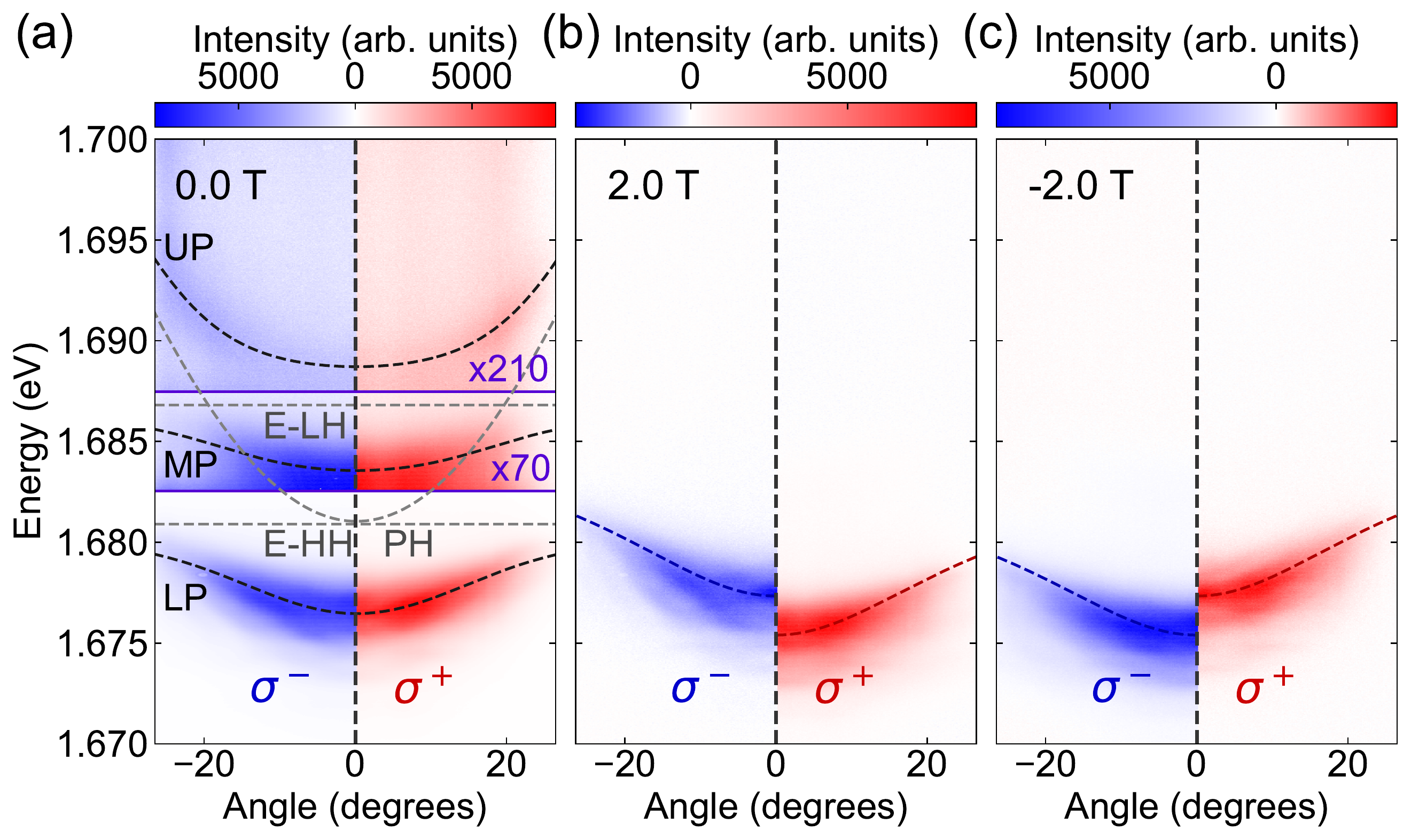}
	\caption{Dispersion of semi-magnetic exciton-polaritons in external magnetic field. Angle-resolved photoluminescence at (a) \SI{0}{\tesla}, (b) \SI{2.0}{\tesla} and (c) \SI{-2.0}{\tesla}. Left/right side of each panel corresponds to $\sigma^-$/$\sigma^+$ polarized emission. Color scale marked by separate colorbars describes emission intensity in $\sigma^-$/$\sigma^+$ polarization with blue/red color. At zero magnetic field fitted dispersion curves (black dashed lines) and obtained bare modes (grey dashed lines) are plotted. For better visibility areas above purple lines are multiplied by indicated factors.}
	\label{fig:fig2}
\end{figure}

The spin properties of exciton-polaritons below and above the condensation threshold in an external magnetic field were studied by angle-resolved photoluminescence. The emission signal below the condensation threshold in $\sigma^-$ (left panel) and $\sigma^+$ (right panel) polarization is illustrated in Fig.~\hyperref[fig:fig2]{\ref{fig:fig2}}. In the absence of a magnetic field [Fig.~\hyperref[fig:fig2]{\ref{fig:fig2}(a)}] we observed equal occupation of $\sigma^+$ and $\sigma^-$ polarized modes. All corresponding counter-polarized states had the same energy. For better visibility, the signal originating from higher modes was enhanced by multiplying by the factors marked in the image. We identified the subsequent states by fitting the three-level model Hamiltonian described in detail in \cite{Mirek_PRB2017}. Dark dashed lines show the positions of the upper (UP) and lower (LP) polariton modes. Gray dashed lines indicate the positions of the heavy hole exciton (E-HH), light hole exciton (E-LH), and photon (PH). Further, in the text, we discuss only the lower polariton branch for two reasons: the occupation of high-energy branches is substantially weaker and the energy splitting in the magnetic field is significantly smaller than the linewidth.

\begin{figure*}
	\centering
	\includegraphics[width=\textwidth]{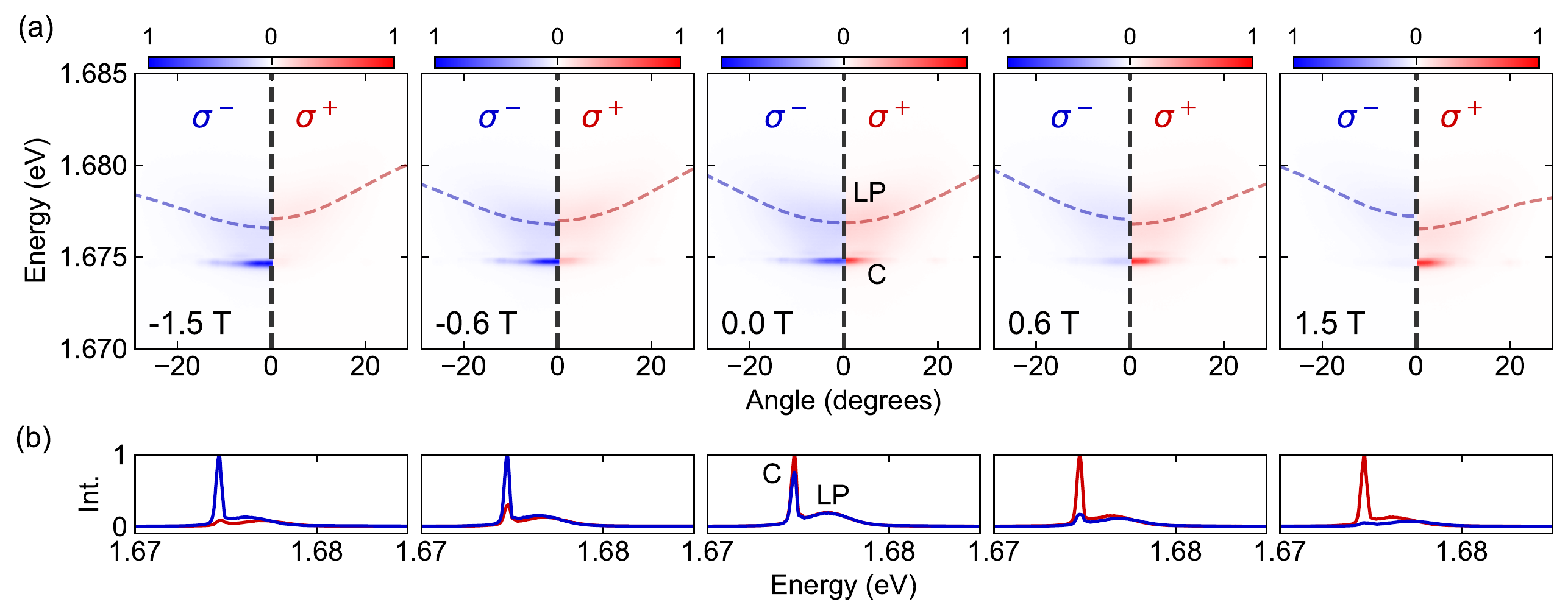}
	\caption{Condensate emission in magnetic field (a) Angle-resolved photoluminescence above condensation threshold (1.5~{$P_{th}$}) in external magnetic field. Left/right side of each panel corresponds to $\sigma^-$/$\sigma^+$ polarized emission. Color scale marked by separate colorbars describes emission intensity in $\sigma^-$/$\sigma^+$ polarization with blue/red color. (b) Emission intensity at zero angle for lower polariton (LP) and condensate (C) in $\sigma^-$ (blue) and $\sigma^+$ (red) polarization. Consecutive panels correspond to the panels above them.}
	\label{fig:fig3}
\end{figure*}

At a magnetic field of \SI{2.0}{\tesla} photoluminescence spectra revealed giant Zeeman splitting [Fig.~\hyperref[fig:fig2]{\ref{fig:fig2}(b)}]. Emission lines in opposite polarizations have different energies resulting from \textit{s,p-d} exchange interaction for excitons. The emission of the $\sigma^+$-splitted branch is more intense than the emission of the $\sigma^-$ polarized LP branch. The difference in population between the two polariton branches mainly results from the Boltzmann distribution. Similarly to 2~T results, the splitting is observed for a negative magnetic field of $-$2~T, as presented in Fig.~\hyperref[fig:fig2]{\ref{fig:fig2}(c)} with reversed occupation factors. In this case, the emission from the $\sigma^-$ polarized branch appears with lower energy and is more intense than the counter-polarized one.

In order to study the spin properties of exciton-polariton condensates, photoluminescence measurements in reciprocal space were performed for excitation power above the condensation threshold. Consecutive panels in Fig.~\hyperref[fig:fig3]{\ref{fig:fig3}} show angle resolved emission at different magnetic fields and cross sections for each polarization obtained from the corresponding maps at $k=0$. In each panel of Fig.~\hyperref[fig:fig3]{\ref{fig:fig3}(a)} two emission lines are visible in both polarizations and originate from the LP and the localized condensate (C). The signal is normalized with respect to the maximum intensity observed in the dominant polarization. At $B=\SI{0}{\tesla}$ we observe strong emission from the condensate and a weak LP signal. The LP branch is linearly polarized (equal population in both circular polarizations) and there is no energy splitting. However, the condensate is elliptically polarized with a dominating $\sigma^+$ component, what is visible in the corresponding cross sections. At \SI{0.6}{\tesla} energy splitting of LP and nonuniform occupation of counter-polarized states is observed. Condensate has the same energy in both polarizations, but the emission in $\sigma^+$ polarization is significantly larger. At a higher magnetic field (\SI{1.5}{\tesla}) even larger Zeeman splitting of the lower polariton is observed, while the condensate does not split in energy due to the spin-Meissner effect \cite{Krol_PRB2019}. In this case right-handed circular polarization is even more dominant both for the condensate and lower polariton. The emission from the condensate is one order of magnitude larger in $\sigma^+$ than $\sigma^-$. Analogous effects are observed at the negative magnetic field. The energy splitting of the lower polariton is inverted and here the $\sigma^-$ polarized branch is dominant. The spin polarization of the condensate is reversed and tends to reach the full occupation of the $\sigma^-$ state with increasing magnetic field.

\begin{figure}
	\centering
	\includegraphics[width=0.48\textwidth]{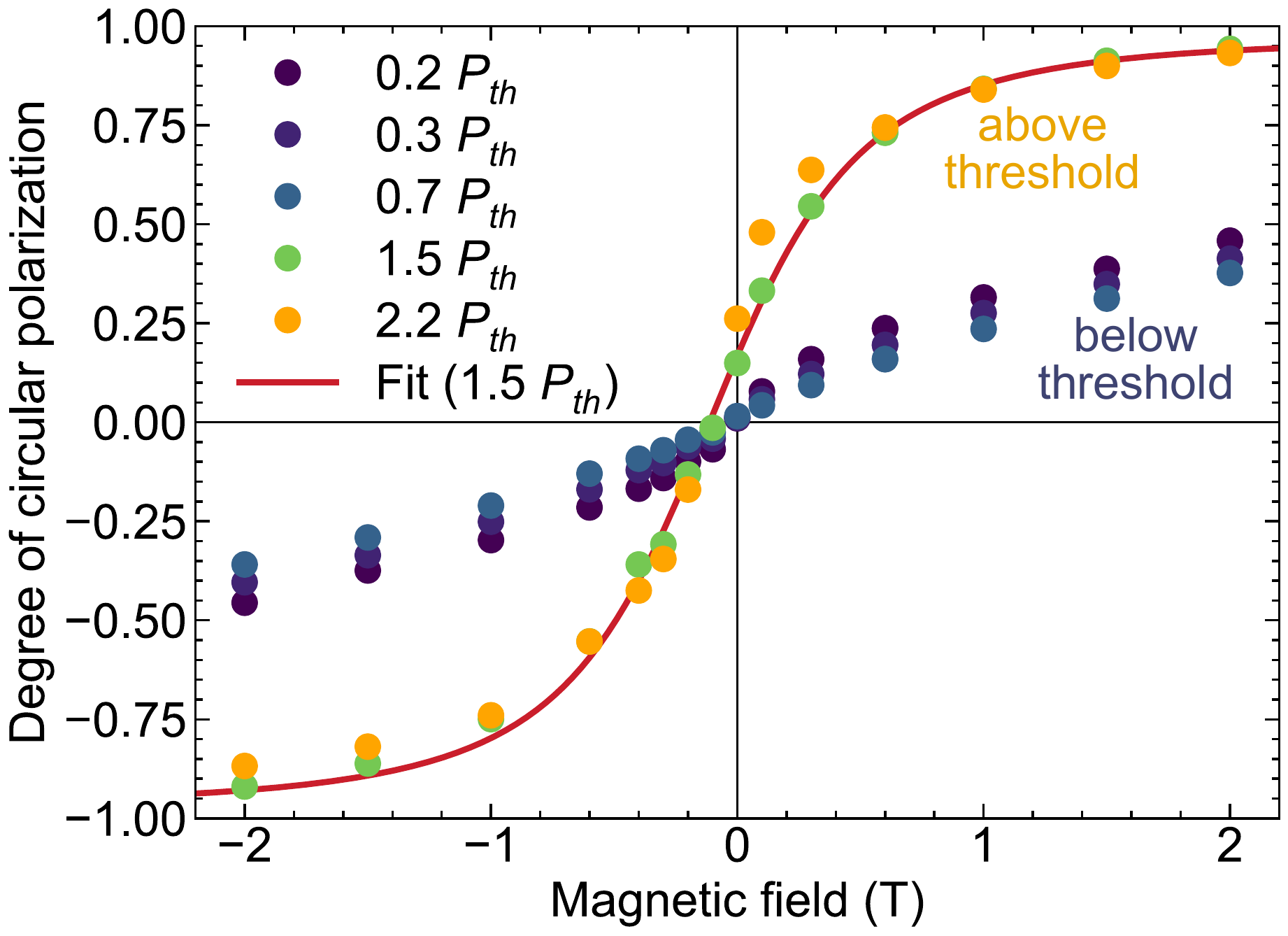}
	\caption{Magnetic field dependence of $\wp$ for different excitation powers with respect to the condensation threshold value, $P_{th}$. The red curve is a fit of Eq.~\eqref{eqn:DOCPShelykhModel} to data for 1.5\,$P_{th}$.
	}
	\label{fig:fig5}
\end{figure}

\begin{figure*}
	\centering
	\includegraphics[width=\textwidth]{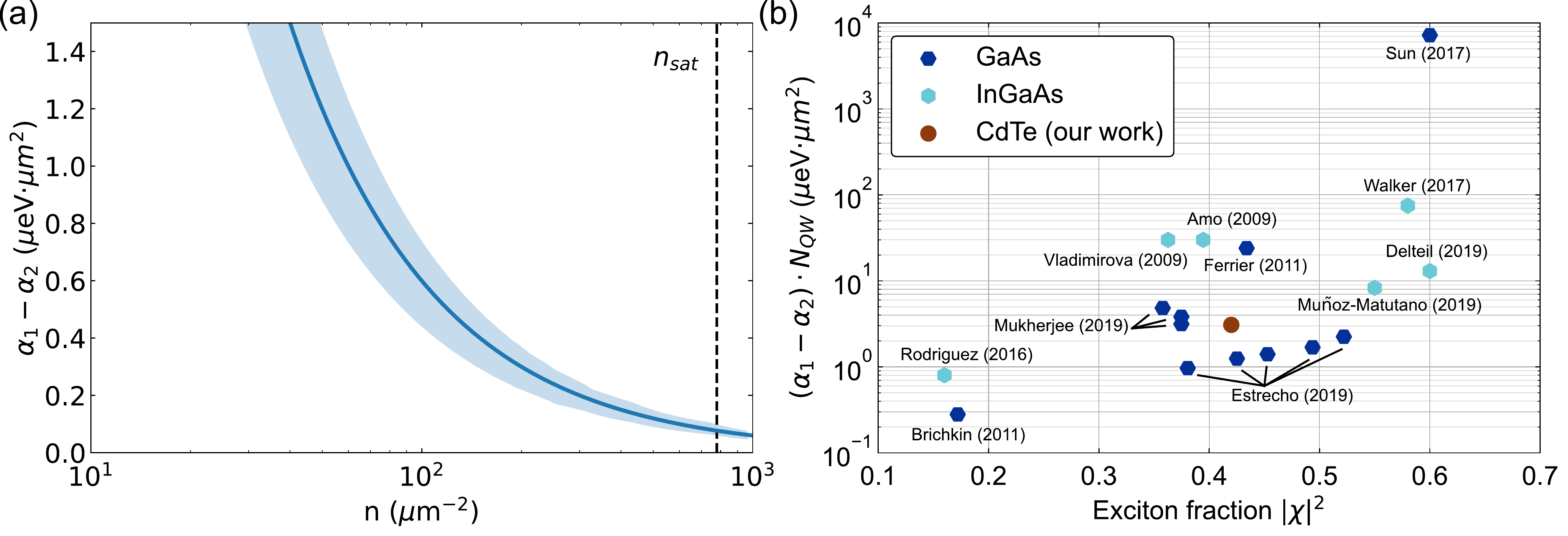}
	\caption{Polariton-polariton interaction strength. (a) Expected polariton strength and polariton concentration obtained from fitting procedure. Blue line corresponds to $n(\alpha_1-\alpha_2)$ = \SI{0.06}{\milli\electronvolt}. The light blue area shows expected error given by uncertainty of manganese ions concentration $n_\text{Mn}$=(1$\pm$0.3)$\%$. The dashed line shows exciton saturation density in CdTe. (b) Previously reported polariton-polariton interaction strength in different systems, based on E. Estrecho et al. \cite{Estrecho_PRB2019}.}
	\label{fig:fig4g}
\end{figure*}

Reciprocal-space measurements were used to extract the magnetic field dependence of the degree of circular polarization of exciton-polaritons, what is illustrated in Fig.~\hyperref[fig:fig5]{\ref{fig:fig5}}. We observe two distinct characteristics for polaritons below and above the condensation threshold. For the excitation power below the condensation threshold, the degree of circular polarization at zero magnetic field is equal to zero. The increase in the magnetic field results in the increase of circular polarization. This behavior is symmetric with respect to the magnetic field. Note that in this case $\wp$ is a result of the imbalance of occupation on both circularly polarized states. The external magnetic field splits the lower polariton branch into two circularly polarized components. Increasing the excitation power in the linear regime slightly decreases the circular polarization due to the heating of manganese ions. Interestingly, above the threshold, the observed relation has a qualitatively different shape. At zero magnetic field condensate manifests nonzero circular polarization (dependent on position on the sample). To compensate this effect, an external magnetic field has to be applied. Moreover, the condensate is more sensitive to the magnetic field and its $\wp$ increases with the magnetic field faster than in the linear regime. The control of polarization is possible because of the presence of magnetic ions in the quantum wells.

To describe the observed degree of circular polarization for the condensate $\wp$ we consider two separate subsystems in the presence of an effective magnetic field: polariton condensate and magnetic ions. We assume that the condensate minimizes its free energy and that the spins of the manganese ions are in equilibrium. This gives us a model of spin polarization of the semimagnetic condensate in the presence of magnetic ions introduced in Ref.~\cite{Shelykh_PRB2009} and modified by an additional term $B_\mathit{eff}$, describing the photonic synthetic magnetic field: 

\begin{equation}
\wp = \frac{n_Mg_M\mu_B\lambda_MW\!\left(\frac{g_M\mu_B(B+B_\mathit{eff})}{k_B\cdot T}\right)}{n\cdot (\alpha_1-\alpha_2)}.
\label{eqn:DOCPShelykhModel}
\end{equation}
Here, \textrm{$n_M$} is a 2D concentration of magnetic ions, \textrm{$g_M$} is a g-factor, \textrm{$\alpha_1$} and \textrm{$\alpha_2$} are the polariton-polariton interaction constants of parallel and anti-parallel spin configuration, respectively, \textrm{$\lambda_M$} is polariton-magnetic ion coupling constant and \textrm{$n$} is polariton concentration. Expression \( W(x) = \sum_{j=-5/2}^{j=+5/2}\frac{j\exp(jx)}{Z(x)}\), where \( Z(x) = \sum_{j=-5/2}^{j=+5/2}\exp(jx)\) is a statistical sum. To take into account the initial polarization observed in the absence of magnetic field we introduced an effective magnetic field $B_\mathit{eff}$ induced by the internal properties of the cavity. Table~\hyperref[tab:DOCPShelykhModel]{\ref{tab:DOCPShelykhModel}} shows the parameters obtained from the fitting procedure together with the calculated values and proposed in Ref.~\cite{Shelykh_PRB2009}.
The obtained g-factor is in very good agreement with g-factor accessed from the giant Zeeman splitting \cite{Mirek_PRB2017}. At \SI{5}{\tesla}:

\begin{equation}
g_M = \frac{\Delta E}{\mu_BB} = \frac{\SI{2.8}{\milli\electronvolt}}{\SI{0.29}{\milli\electronvolt}}=9.7. 
\label{eqn:gfactor}
\end{equation}

\begin{table}
\caption{\label{tab:DOCPShelykhModel} Parameters obtained from the fitting procedure (Fit) and calculations (Calc.) compared with parameters assumed in Ref.~\cite{Shelykh_PRB2009}.}
\begin{tabular}{l l l l}
    \hline \hline
	\textbf{Parameter} & \textbf{Fit}  & \textbf{Calc.} & \textbf{Ref.~\cite{Shelykh_PRB2009}}\\
	\hline
	$n_M$ [mm$^{-2}$]& 5.8$\cdot10^{10}$ & \textendash  & 5$\cdot10^{12}$\\
	$g_M$ & \textendash & 9.7 & 2.02\\
	$n(\alpha_1-\alpha_2)$ [meV] & 0.06 & \textendash & 2.4\\
	$\lambda_M$ [T$\cdot$mm$^2$] & \textendash & 7.0$\cdot10^{-13}$ & 2.56$\cdot10^{-11}$\\
	$B_\mathit{eff}$ [T] & 0.11 & \textendash & \textendash\\
	$T$ [K] & 5.0 & \textendash & \textendash\\
\end{tabular}	
\end{table}

The coupling constant between polariton and magnetic ion can be evaluated following:
\begin{multline}
\lambda_M = \frac{\beta_{exc}\chi^2}{\mu_Bg_ML_{QW}N_{QW}}=\\
=\frac{\SI{1.1}{\electronvolt}\cdot\frac{\left(0.648\,\text{nm}\right)^3}{4}\cdot0.42}{\mu_B\cdot 9.7\cdot 20\,\text{nm}\cdot 4}=7.0\cdot 10^{-13}\,\text{T}\cdot \text{mm}^2,
\label{eqn:lM}
\end{multline}
\begin{figure*}
	\centering
	\includegraphics[width=\textwidth]{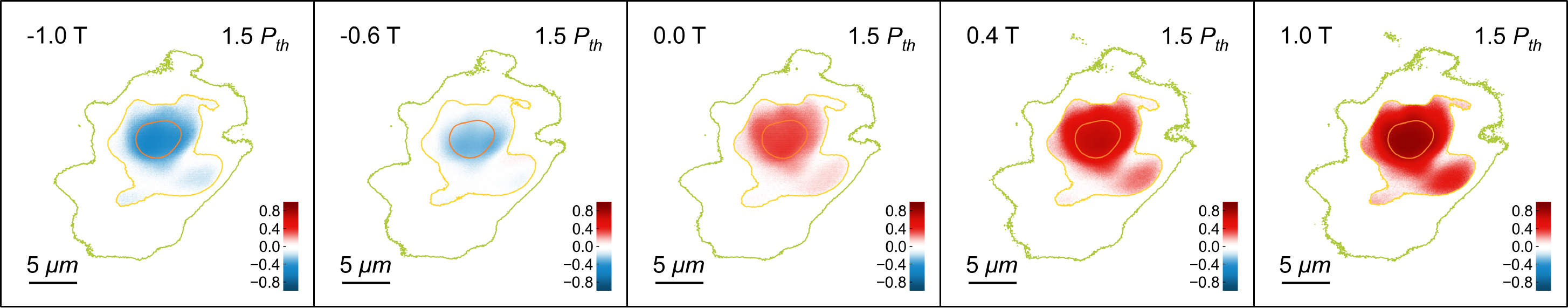}
	\caption{Degree of circular polarization of exciton-polariton condensate studied at different external magnetic field in real space. Magnetic field, excitation power, and scale are indicated in the annotations.}
	\label{fig:fig4}
\end{figure*}
where $\beta_{exc}$ origins from the exchange integrals of CdMnTe \cite{Gaj_SSC1979} $\left(\beta_{exc} = \frac{\SI{1.1}{\electronvolt}}{N_0}\right)$ and CdMnTe cation concentration $\left(N_0 = \frac{4}{(0.648\,\text{nm})^3}\right)$. Here, $\chi^2$ is an excitonic Hopfield coefficient and $L_{QW}$ and $N_{QW}$ are width and number of quantum wells, respectively. 

Comparing all the values given in Table~\hyperref[tab:DOCPShelykhModel]{\ref{tab:DOCPShelykhModel}}, we find a relatively good agreement either with our calculated values or previously estimated for the CdMnTe-based cavity. From the two-dimensional concentration of manganese ions, we calculated the manganese ions concentration $n_\text{Mn}$:

\begin{equation}
n_\text{Mn} = \frac{n_M}{N_0^\frac{2}{3}} = 1.0\%.  
\label{eqn:nMn}
\end{equation}

Particularly important is the estimation of the polariton-polariton interaction constant. From the curve fitting procedure, $n(\alpha_1-\alpha_2)$ of \SI{0.06}{\milli\electronvolt} was obtained. Estimation of the polariton-polariton interaction constant is based on Fig.~\hyperref[fig:fig4g]{\ref{fig:fig4g}(a)} where we plotted the expected interaction constant for a given polariton concentration. The solid blue line corresponds to the value obtained from fitting. The expected concentration must be lower than the exciton saturation density marked with a dashed line \cite{Huang_PRB2002}. We assume the polariton concentration to be one order of magnitude below this limit \cite{Kasprzak_PRL2008} what gives a rough estimation of the polariton interaction strength of about 0.8\,$\mu \text{eV} \!\cdot \!\mu \text{m}^2$. We compare in Fig.~\hyperref[fig:fig4g]{\ref{fig:fig4g}(b)} deduced polariton interaction strength with the values obtained for GaAs \cite{Brichkin_PRB2011, Ferrier_PRL2011, Sun_PRL2017,Estrecho_PRB2019, Mukherjee_PRB2019} and InGaAs \cite{Amo_NatPhys2009, Vladimirova_PRB2009, Rodriguez_NatComm2016, Walker_PRL2017, Delteil_NatMater2019, MunozMatutano_NatMater2019} quantum wells, previously summarized in Ref. \cite{Estrecho_PRB2019}. CdTe-based structures have higher exciton binding energy, but smaller Bohr radius than GaAs-based structures, so the value of interaction constant in the triplet spin configuration is expected to be similar in both materials. It is worth noting that our measurements were performed under nonresonant excitation, what leads to creation of an excitonic reservoir. The main contribution from excitonic reservoir is the effect of blueshift of the condensate energy, what makes difficult to estimate the interaction constant from condensate energy shift.  

The effect of synthetic magnetic field present in exciton-polariton condensates can also be observed in real space. Fig.~\hyperref[fig:fig4]{\ref{fig:fig4}} illustrates the spatial extent of the degree of circular polarization of a localized condensate in different external magnetic fields, for the excitation power above the condensation threshold (1.5\,$P_{th}$). At high magnetic field (of \SI{1}{\tesla}) condensate is almost fully circularly ($\sigma^+$) polarized. The decrease in the magnetic field results in a decrease of $\wp$ of the condensate. However, at zero magnetic field nonzero degree of circular polarization is observed. By using a high enough external magnetic field in the opposite direction, it is possible to reverse the spin of the condensate. For \SI{-0.6}{\tesla} $\wp$ of condensate is negative. A further increase of the magnetic field results in a stronger build-up of $\sigma^-$ polarization. As a side note, we remind that the initial degree of circular polarization in the absence of magnetic field depends strongly on position on the sample and can be both positive and negative. Both the sign and value of $\wp$ vary significantly while changing the position of the linearly polarized excitation laser spot.

\subsection{\label{sec:results_C} Spin of the condensate tuned by the polarization of non-resonant laser}

\begin{figure*}
	\centering
	\includegraphics[width=\textwidth]{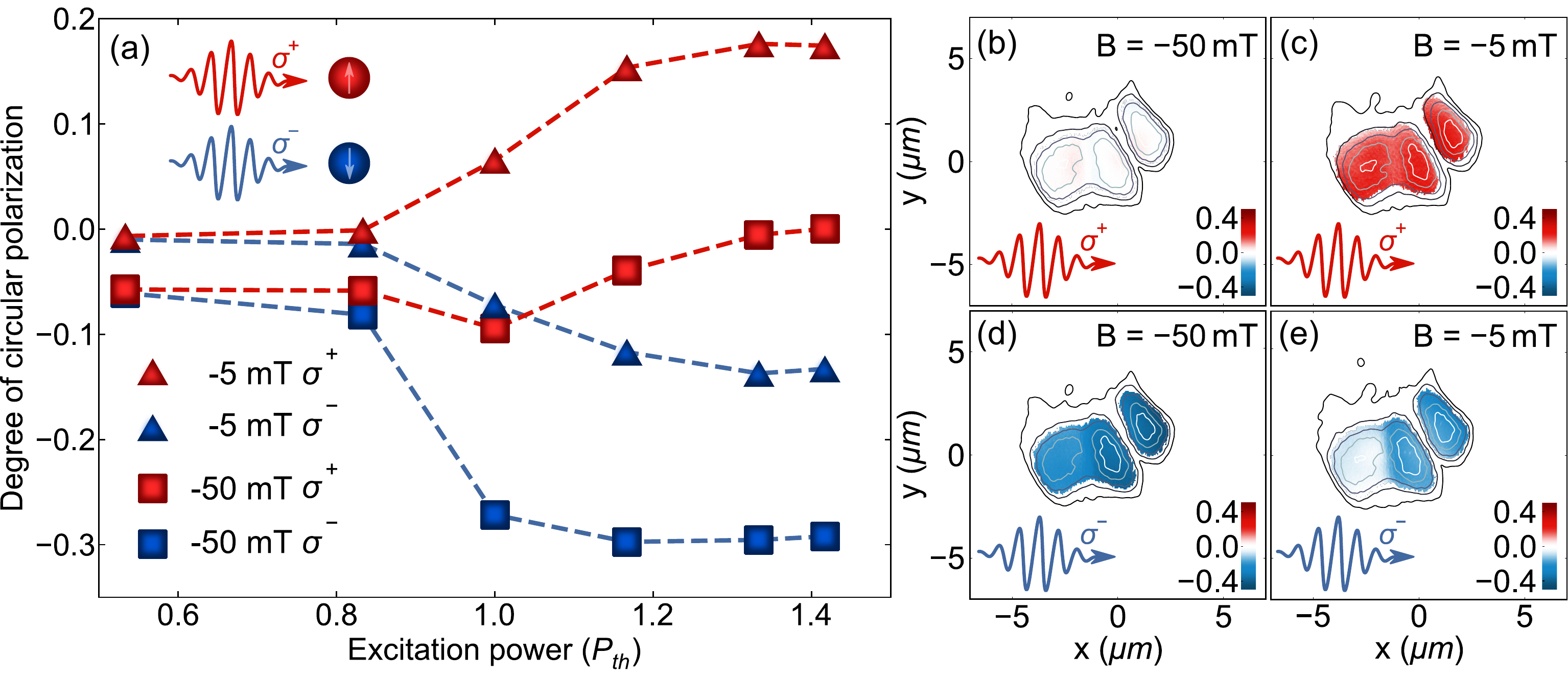}
	\caption{Influence of laser polarization on $\wp$. (a) Power dependence of $\wp$ for circularly polarized laser at \SI{-5}{\milli\tesla} and \SI{-50}{\milli\tesla}. Legend placed in left bottom corner indicates external magnetic field and polarization of excitation laser. Inset in top left corner shows brief scheme of the experiment. (b)--(e) Degree of circular polarization in real space for excitation power above threshold ($1.42\,P_{th}$). Magnetic field (/polarization of the laser) is marked in a top right (/bottom left) corner.}
	\label{fig:fig6}
\end{figure*}

All previous results were performed for linear excitation of the laser. Here, we show the influence of the circularly polarized laser on the degree of circular polarization of exciton-polaritons for different excitation power  (Fig.~\hyperref[fig:fig6]{\ref{fig:fig6}}). Below the condensation threshold, the polarization was mostly affected by the external magnetic field. Even in a linear regime, the spin polarization of polaritons was slightly changing with the polarization of the external laser and this effect was pronounced for higher excitation power. The condensate was even more sensitive to the polarization of the external laser. Above the condensation threshold, an increase of excitation power of $\sigma^+$(/$\sigma^-$) polarized laser led to increase(/decrease) of $\wp$. The influence of the circularly polarized pump was visible even in the external magnetic field. The real space images of the degree of circular polarization for three condensates pumped with the excitation power of 1.42\,$P_{th}$ are presented in [Fig.~\hyperref[fig:fig6]{\ref{fig:fig6}(b-e)}]. At the magnetic field of \SI{-50}{\milli\tesla}, pumping the condensates with $\sigma^+$ polarized laser resulted in almost zero degree of circular polarization. The change of the laser polarization into $\sigma^-$ was followed by the appearance of the evident $\sigma^-$ polarization component. 
At the lower magnetic field (of \SI{-5}{\milli\tesla}), condensates were elliptically polarized with dominating $\sigma^+$ component for $\sigma^+$ excitation. For the opposite polarization of the excitation laser, we observed a change of the sign of the degree of circular polarization of the condensates.

\section{Summary}	
In this paper we created localized condensates with elliptical spin polarization being the consequence of the inhomogeneous photonic potential on the sample. The spin polarization of the condensate depends on the photonic potential, therefore on the position on the sample and results from the nonideal growth of a distributed Bragg reflectors and microcavity. The presence of the disordered potential has its crucial role in lifting the spin degeneracy of the condensate and leads to the observation of a nonzero degree of circular polarization of the emission from the condensate in the absence of an external magnetic field. Our results demonstrate that in semiconductor microcavity, the time-reversal symmetry can be intrinsically broken. This gives the opportunity to create arrays of condensates with different polarization patterns using a linearly polarized laser beam. 
We introduced also a method to assess polariton-polariton interactions in semimagnetic quantum wells from a build-up of condensate spin polarization in an external magnetic field. We estimated the interaction strength of 0.8\,$\mu \text{eV} \!\cdot\! \mu \text{m}^2$ which is in good agreement with the values previously published for other materials. Even though this method is restricted to semimagnetic materials it is based on magnetic field measurements and is complementary to previously reported. 

\section{Acknowledgements}
This work was supported by the National Science Center, Poland, under the projects 2019/33/N/ST3/02019 (R. M.), 2015/18/E/ST3/00558 (B. P. and K. T.), 2017/27/B/ST3/00271 (K. Ł. M.), 2016/22/E/ST3/00045 (M. M.), and 2020/37/B/ST3/01657 (M. F.). This study was carried out using CePT, CeZaMat, and NLTK infrastructures financed by the European Union, European Regional Development Fund.

\bibliography{bibliography}

\end{document}